\documentclass[journal]{IEEEtran}
\usepackage[utf8]{inputenc}

\usepackage{amsmath}
\usepackage{siunitx}
\usepackage{multicol}

\DeclareMathOperator*{\argmin}{arg\,min}
\DeclareMathOperator{\Tr}{Tr}
\usepackage{graphicx}

\usepackage{txfonts}
\usepackage{relsize}
\usepackage{subfig}
\usepackage[numbers,compress]{natbib}

\ifCLASSINFOpdf
\else
\fi

\usepackage{stfloats}
\begin{document}
%
\title{Signal estimation and uncertainties extraction in TeraHertz Time Domain Spectroscopy}
%
%
%

\author{ \IEEEauthorblockN{Elsa Denakpo\IEEEauthorrefmark{1}, Théo Hannotte\IEEEauthorrefmark{1}, Noureddin Osseiran\IEEEauthorrefmark{1}, François Orieux\IEEEauthorrefmark{2}, and Romain Peretti\IEEEauthorrefmark{1}}
\thanks{The authors marked with \IEEEauthorrefmark{1} are with Univ. Lille, CNRS, Univ. Polytechnique Hauts-de-France, UMR 8520 IEMN - Institut d'Electronique de Microélectronique et de Nanotechnologie, F-59000 Lille, France, marked with \IEEEauthorrefmark{2} with Université Paris-Saclay, CNRS, CentraleSupélec, Laboratoire des signaux et systèmes, 91190, Gif-sur-Yvette, France. Correspondance should be sent to romain.peretti@cnrs.fr}}

%
%

\markboth{Journal of \LaTeX\ Class Files,~Vol.~14, No.~8, August~2015}%
{Shell \MakeLowercase{\textit{et al.}}: Bare Demo of IEEEtran.cls for IEEE Journals}
%



\maketitle

\begin{abstract}
Terahertz Time Domain Spectroscopy (THz-TDS) systems have emerged as mature technologies with significant potential across various research fields and industries. However, the lack of standardized methods for signal and noise estimation and reduction hinders its full potential. This paper introduces a methodology to significantly reduce noise in THz-TDS time traces, providing a reliable and less biased estimation of the signal. The method results in an improved signal-to-noise ratio, enabling the utilization of the full dynamic range of such setups. Additionally, we investigate the estimation of the covariance matrix to quantify the uncertainties associated with the signal estimator. This matrix is essential for extracting accurate material parameters by normalizing the error function in the fitting process. Our approach addresses practical scenarios where the number of repeated measurements is limited compared to the sampling time axis length. We envision this work as the initial step toward standardizing THz-TDS data processing. We believe it will foster collaboration between the THz and signal processing communities, leading to the development of more sophisticated methods to tackle new challenges introduced by novel setups based on optoelectronic devices and dual-comb spectroscopy.
  
\end{abstract}



%
\IEEEpeerreviewmaketitle

\section{Introduction}
Terahertz (THz) frequency range has long been a frontier for technology, too fast for electronics and at too low energy to ensure proper laser radiation. Therefore, THz spectroscopy was only rarely used and only for research purposes despite much interest in gas, liquid, and solid phase samples. The paradigm shifted a few decades ago thanks to the use of opto-electronics sources coupled to new generations of both continuous and pulse lasers. This enabled the performances of THz spectroscopy from a few fraction of THz to a few THz with a very good dynamic range using a bench-top setup. This highly contributed to the spread of THz spectroscopy, which is used to characterize the physical and chemical properties of samples from the solid, liquid, gaseous, and even plasma phases \cite{houver2023}.

In gas, THz covers both the rotational transitions of the microwave regime for relatively small molecules (3 to 15 atoms) and the vibrational modes of the infrared for larger molecules. Therefore, it is known to be the most selective spectral range for organic volatile compounds \cite{Smith2015} with potential application in atmospheric study, air pollution monitoring, or breathalyzers. In liquids, water picosecond dynamics is in the same time an open fundamental subject and very sensitive to interaction with solutes and thus provides important information about proteins solvation \cite{adams2020, cheng2020} for instance. Finally, in the solid state, it is used in many fundamental semiconductor \cite{failla2016,biasco2022,rajabali2022}, spintronic \cite{dang2020ultrafast,seifert2018}, and 2D material research \cite{valmorra2013,deinert2020}, as a very good tool to monitor the crystal phase of organic molecules \cite{mitryukovskiy2022,bawuah2021}. THz is therefore very attractive for many analytical fields and is used to characterize tablets porosity during their fabrication in the pharmaceutical industry and in the automotive one - with the major example of measuring the thickness of paint layers during the painting of automotive parts or non-destructive analysis of thin film elements.

The setups and techniques have developed now and commercially available time-domain spectroscopy setups offer more than \SI{100}{dB} of dynamics range and bandwidth of about \SI{8}{THz} and are distributed and used by scientists and engineers all over the world. THz spectroscopy is a rapidly growing field with a wide range of applications. During the experiments the instrument records two time traces one without sample and one with sample. From there, several teams have proposed methodologies to extract useful information for the physicist, chemist or material scientist. Historically \cite{duvillaret1996reliable,duvillaret1999,jiang2000,dorney2001}, the sequence of information extraction included, a FFT, phase unwrap, refractive index evaluation thanks to a basic and often valid photonic model, and finally the parameter estimation. They used either a better model for the recorded signal \cite{tayvah2021nelly}, or went further in classifying samples through data processing \cite{liao2023amino, WANG2022}. The last approach was to present statistically valid estimation to extract the parameters when fitting the experimental data \cite{Mohtashemi2021}. Still, despite these efforts \cite{lee2023}, there is no standard in the data acquisition and data processing and no accepted protocol to extract the uncertainties from a measurement and therefore rare are the studies proposing error bars in their measurements.

In this work, we propose to go a step further in the data acquisition and processing in THz-TDS experiments thanks to a robust estimator of the signal on the one hand, and an estimation of the noise correlation matrix needed for materials parameters extraction on the other hand. The estimator is based on the average of the time traces corrected depending on the source of noise, by applying a generic optimization approach. The noise correlation matrix for its part is approximate with algorithms used for covariance and precision matrices estimation. Both are included in the open source free software Correct@TDS \cite{correct@tds}, which implements our estimator and enables the estimation of the noise correlation matrix. The manuscript first shortly describes the THz-TDS setup and the methods used in the community to process the data, specifies the different sources of noise and ways to mitigate it, proposes an estimator for the signal and the corresponding noise correlation matrix. We believe that this approach and software will help the technique to spread more in analytical fields and hopefully to draw a bridge to the data processing and instrumentation community.
\section{TeraHertz Time Domain spectroscopy setup, data exploitation and experimental errors}

\subsection{TeraHertz Time Domain spectroscopy experiments}
   \begin{figure}[!htb]
   \centering
   \includegraphics[width=6cm]{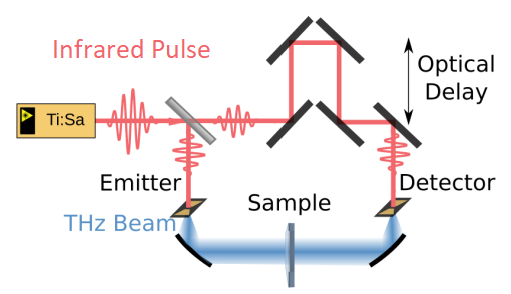}
   \caption{THz-TDS experiment principle showing the main elements and source of errors through noises and perturbations: the delay line and the detector and the laser.}
              \label{figtds}
    \end{figure}
In a THz-TDS experiment (Fig.~\ref{figtds}), a femtosecond laser emits an ultrafast pulse that is split in two by a beam splitter and directed to two photoconductive antennas. When the ultrashort pulse excites the transmitting antenna, it creates carriers which are accelerated by an high voltage between two electrodes and induce the terahertz pulse. The detectors works on a very similar principle and the delay line allows to adjust the length of the path before reaching the detection antenna. Thus by varying the time delay between the pulse on the detector and the one on the emitter, the E-field is measured as a function of time. At the end of an experiment, we obtain two time traces, one without sample that we call the reference and one with sample that we call sample. In both cases, the electric field measured versus time, which will be analyzed to retrieve information on the properties of the studied material.
\subsection{TeraHertz Time Domain spectroscopy First data process}
From theses time traces, several further steps have been proposed to extract information from the experiments. The first one is the computing of the Transmission; it consists on a simple performance of the ratio between the absolute value of the FFT of the time traces. The second one add a log to the previous step to get the loss per unit length when divided by the thickness of the sample. The third one is a bit more complex,  it consists on the extraction of the complex refractive indices versus frequency, of the studied material. It is again usually done from the frequency domain by Fourier-transform of the recorded time traces. Then, a fitting process is used, to get the two real value describing the complex refractive index \cite{herault2015possibility}. Finally, one can further extract information by fitting the dispersion of the refractive index by permittivity models to get physical insight about the material.
This method in the frequency domain reach its limitations for instance when the signal is too noisy due to a too strong absorption \cite{bernier2016}. Indeed, the accuracy of the measured time traces have a direct influence on the performance of the fitting model. Moreover, the fit by the permittivity models usually do not take into account the noise. This omission would be valid if the noise would be a Gaussian white noise but in the experiments this assumption was actually never confirmed. To overcome this issue, several group decided to look back to the initial time traces in the performance of the data processing. \citeauthor{mohtashemi2021maximum} \cite{mohtashemi2021maximum} proposed a method that includes the noise correlation matrix which is a fair description of the uncertainty due to the noise in the reference and in the sample time traces. In our group \cite{peretti2018thz} and \cite{lavancier2021criterion, lavancier2024}, we proposed a software called Fit@TDS which implements a method where the permittivity models are directly fitted from the data in time domain and include the noise correlation matrix.
This two new method show the importance of looking deeper on the source of error from the systems in THz-TDS experiments.
\subsection{THz-TDS systems noises and perturbations}
THz-TDS systems are now well matured but their capabilities can still be improved beside simple the output power of the emission antenna and the bandwidth of the systems. Indeed, the interpretability and the reproductibility of the output data is of utter importance specifically regarding application such as in biology \cite{markelz2022}. This relies on a deep understanding of all the sources of error, including noises and perturbations, induce by the system itself. THz-TDS are known to have a very good dynamic range, but up to now there is no standard definition of the signal to noise ratio. Thus it's important to understand where the noise comes from the system.  Many studies \cite{duvillaret2000,Withayachumnankul2008,naftaly2012,muller2021,lavancier2021thesis,denakpo2022noise} had been done on the subject of noise sources in THz-TDS systems. \citeauthor{jahn2016influence} \cite{jahn2016influence} described how the uncertainties in the delay line positioning influences the acquired signal and its spectrum. \citeauthor{rehn2017periodic} \cite{rehn2017periodic} explained how the periodic error in delay line line position can make the spectrum unusable. Hence, the main sources of noise identified are the variation of the laser power and the delay line position which impact the time sampling step. 

To better quantify and understand the noise added to the signal by the spectrometer, we performed a set of  experiments. Our setup is a MenloSystems's terahertz spectrometer TeraSmart. This setup is relatively standard using a \SI{100}{MHz} repetition rate laser and a \SI{850}{ps} long delay line covering 8.5\% of the laser period with a time step of \SI{33}{fs}. The typical THz bandwidth is around \SI{4}{THz}.

For the sake of comprehension of the rest of this work, we denote:
\begin{gather}
    T_i = \left(E_i(t)\right)_h\\
    \tilde{T}_i = \left(\tilde{E}_i(f)\right)_{\ f  \in \left\{0, \cdots, \left\lfloor \frac{p-1}{2}\right\rfloor\right\}}\\
    \mathcal{T} = \left(T_i\right)_{\ i\  \in \left\{1, \cdots, n\right\}}\\
    \overline{T} = \left(\frac{1}{n}\sum_{i = 1}^{n} E_i(t)\right)_{\ t\  \in \left\{t_0, \cdots, t_{p-1}\right\}}
\end{gather}

\noindent with $T_i$ the $i$th measured time trace, $\tilde{T}_i$ its Fourier transform, $\mathcal{T}$ the set of $n$ time traces measured repeatedly, and $\overline{T}$ the average along the time axis of all time traces $T_i, i \in \{1 \cdots n\}$.

\begin{figure}[!htb]   
   \centering
   \includegraphics[width=8cm]{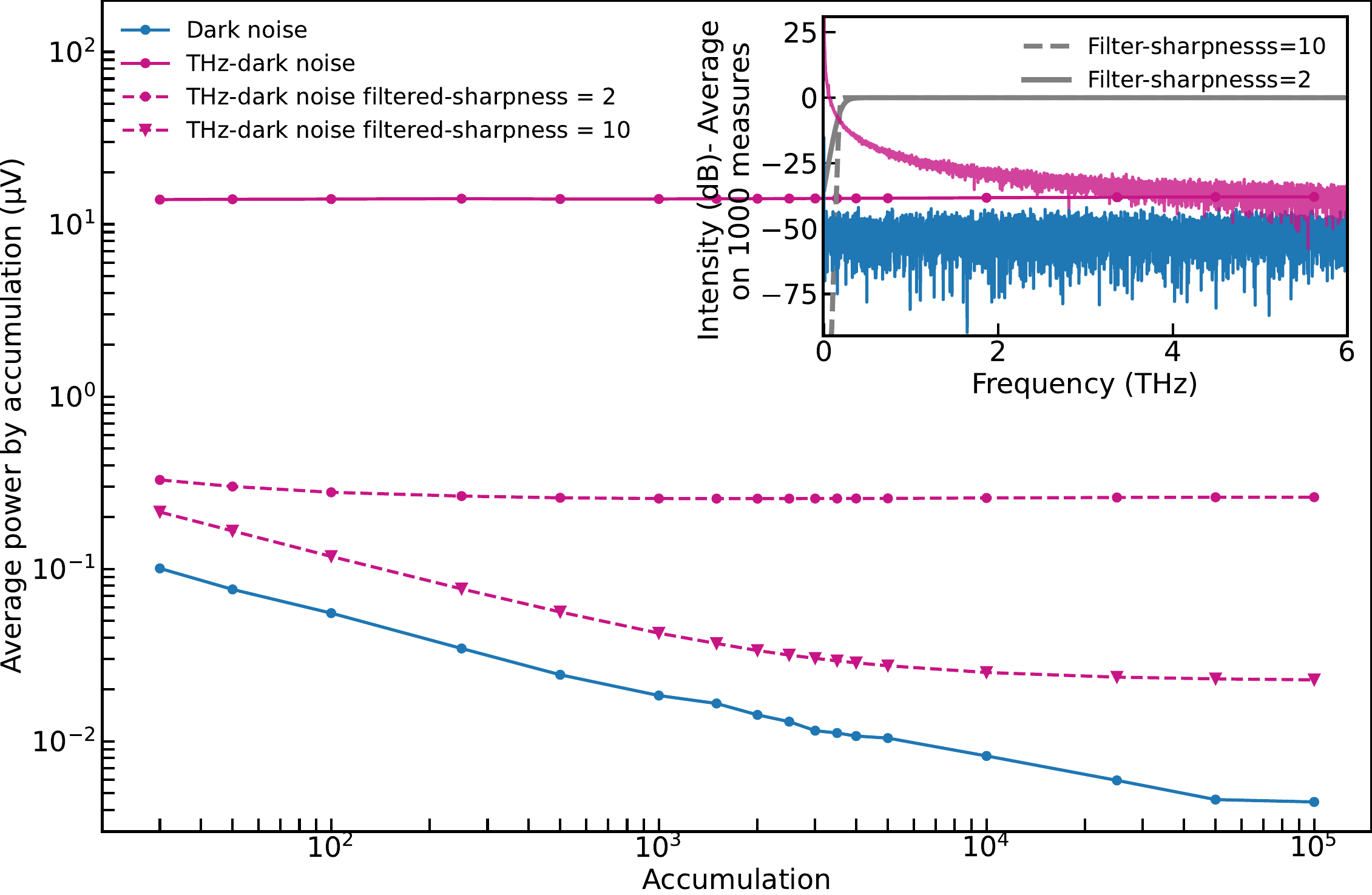}
   \caption{Noise recorded in the different experimental conditions: Dark noise and THz-Dark noise with different filtering showing that this filtering is unavoidable to keep a reasonable noise even in the absence of signal. }
              \label{fignoise}
    \end{figure}

For the first experiment the two optical fibers are unplugged and we recorded the signal on \SI{845}{ps} without THz emission and activation of the detector. We call this noise "Dark noise" and it is generated by the detector when no THz pulse is sent to the detector. In fact, even when no carrier are promoted to the conduction band, some hole-electron pairs are created, generating a small current, that is amplified by the transimpendance amplifier, itself adding noise. Fig.~\ref{fignoise} shows the average on \num{1000} single record in frequency domain. Its average power decreases when accumulating, it goes from \SI{1e-1}{\micro\volt} for an average on $30$ time traces to \SI{1.1e-2}{\micro\volt} for an average of \num{3000} time traces. It is non deterministic and has the flat spectrum of a white noise.

A second experiment is to put an absorbent to block the cell occupied by the sample and record the signal on \SI{845}{ps}. Both  antennas are excited by the femtosecond laser and we call this "THz-Dark perturbation"; its amplitude as well as its shape include a low frequency artifact-perturbation that is reproducible \cite{lavancier2024}. As shown in Fig.~\ref{fignoise}, its amplitude is two orders of magnitude greater than the Dark noise recorded previously in our first experiment and its shape has a deterministic trend. Here, we thoroughly ensure that this noise was not due to parasitic THz reflection in our setup by translating and rotating the antenna without changing this perturbation. This implies that some parasitic laser emission reach the detector after the main pulse and after going through the delay line. At low frequency, its contribution is important and must be reduced. This was done thanks to a low cut filter at \SI{200}{GHz} corresponding to the low bounds frequency of the bandwidth given by the manufacturer of our setup. The filter is a  smooth step function and is defined by:
\begin{equation} \label{eq0}
    H(f) = 0.5+0.5\tanh\left( \frac{(f-f_{cut})\beta}{f_{cut}} \right)
\end{equation} 
with $f_{cut}$ the cut-off frequency and $\beta$ the sharpness. Fig.~\ref{fignoise} shows the average power of this noise. After filtering for an accumulation of \num{3000} time traces the THz-Dark noise average power goes from \SI{14}{\micro\volt} before filtering to \SI{3e-2}{\micro\volt} after filtering for a sharpness of 10, similar to the Dark noise order of magnitude.
    

From this experiments, it appears that the signal should be filter at low frequencies in order to reduce both noises. Also the sharpness of the filter can have a great influence on the result. In fact, \SI{200}{GHz} is the frequency limit of our system, and a large sharpness means more correlation added to the data. Hence a filter under \SI{200}{GHz} with a sharpness of $2$ is a good compromise to not impact the most important part of the signal at higher frequencies.

\begin{figure}[!htb]
   \centering
   \includegraphics[width=8cm]{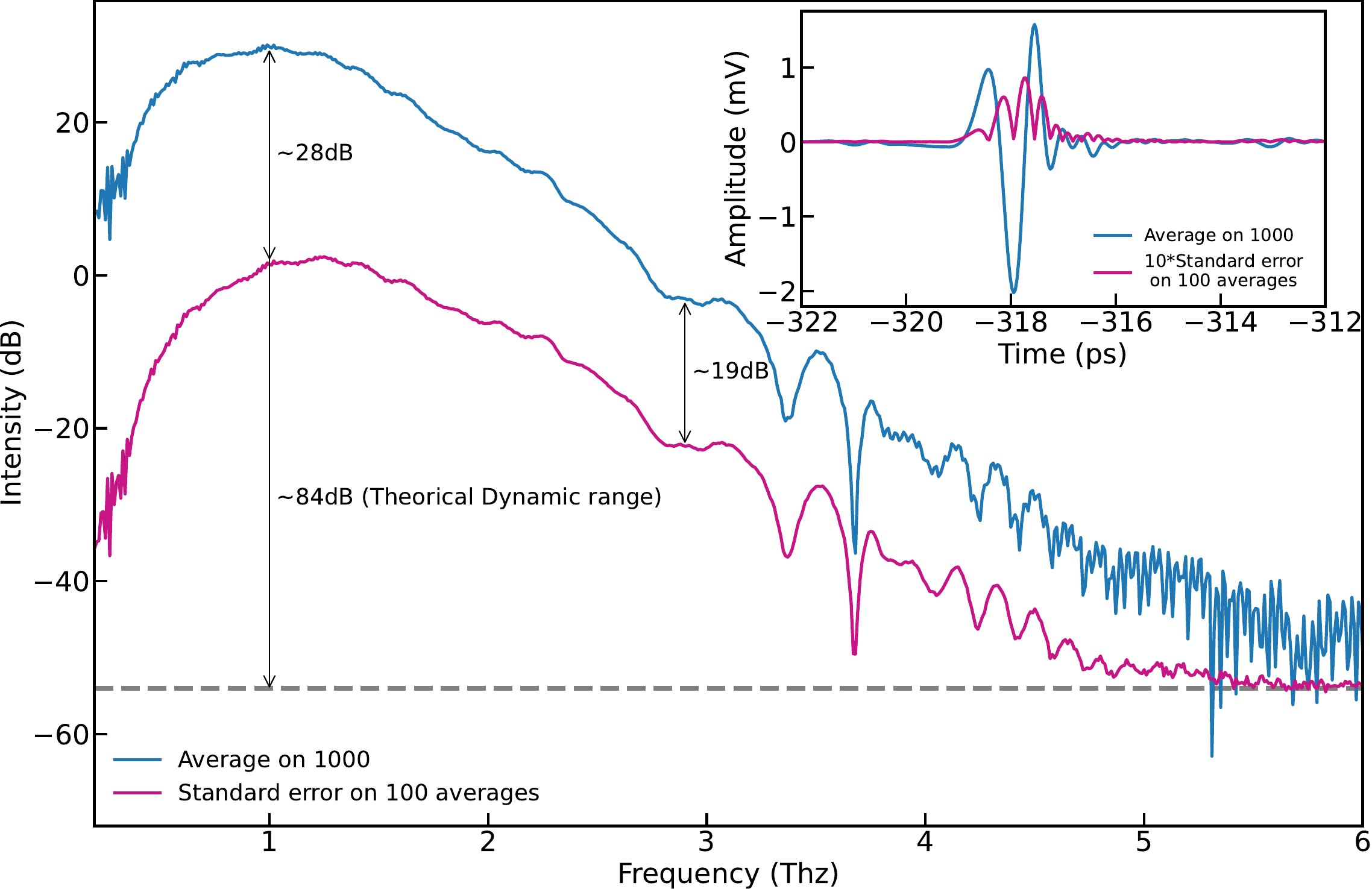}
   \caption{Mean and standard error in frequency and time domain on time traces recorded in dry nitrogen showing that the noise cannot be said to be equal to neither the dynamic range nor to the dark noise.}
              \label{figraw}
    \end{figure}

A third experiment is to record THz pulse traveling in dry nitrogen between the two antennas, plugged to the optical fibers. Here \num{100000} time traces on \SI{100}{ps} have been acquired consecutively. Fig.~\ref{figraw} shows the average on \num{1000} time traces and the standard error on $100$ averages on \num{1000} time traces.

When averaging on time traces samples, the goal is to increase the signal to noise ratio. The average is then considered as an estimator of the signal and the standard error it's standard deviation. Indeed, in statistics, when the samples are independent and identically distributed (i.i.d.) from a Gaussian distribution with mean $\mu$ the average - also called the sample mean - is the maximum likelihood estimator of this mean $\mu$.

As we can see in Fig.~\ref{figraw}, the standard error is correlated to the signal and the signal to noise ratio is about \SI{28}{dB} while the dynamic range as define in the community (maximum of the signal compare to the high frequency noise) is $\sim\SI{84}{dB}$. The first conclusion is that for the estimator "average", the noise estimated by the standard error is very high compare to the dynamic range. Two causes can lead to this effect: There is a very important noise correlated to the signal; the estimator "average" on the raw data is not a good estimator for this experiments. In the next sections, we will propose to correct most of the noise coming from the system to build a better estimator of the signal and increase the signal to noise ratio.

\section{Signal estimator}

\subsection{Delay line initial position drift}

While accumulating multiple measurements is a common approach to improve signal estimation, the simple average often results in a biased estimator for time-domain sampled signals as already noted in \cite{souders1990}. Specific efforts have been pursued to retrieve unbiased signal estimators in time domain experiments. In our case, upon closer examination of the data, we discovered that the time traces exhibited slight shifts of a few femtoseconds between each accumulation. It translates in a shape of the standard deviation of the time signal similar to the one shown in \cite{verbeyst2006} and the inset of \ref{figraw}. These shifts, attributed to temperature variations in optical fibers altering the optical delay or positioning changes in the delay line, lead to misalignment of the main pulse in each time trace. As a result, when these traces are averaged, noise is introduced, as depicted in Fig.~\ref{figraw}. This issue of signal alignment is quite general \cite{coakley2001} and not unique to THz-TDS experiments; similar challenges exist in other fields such as sampling oscilloscopes \cite{verbeyst2006,coakley2001} , seismology \cite{harris1991,romano2016}, radar\cite{wang2023}, and ultrasound for biology \cite{al2020}. While several methods have been proposed to address this problem, we initially chose to explore the simplest solutions.

To correct this misalignment smaller than the size of a time sampling step, we implemented the standard formula of the Fourier transform of a translation. A time trace among the recorded time traces is chosen as a reference and all time traces  are readjusted one by one to this time trace before averaging. The reference time trace is defined as the time trace  closest to the average:

   \begin{equation} \label{eq1}
    T_{ref} = \argmin_{T_i \in \mathcal{T}}\ \frac{\overline{T} \cdot T_i}{||T_i||_2}.
\end{equation} 

By naming $\delta_i$ the delay shift on the time trace $i$, the corrected signal in frequency domain is: 
\begin{gather}
    \tilde{E}_{i-corrected}(f) = \exp{(j. 2\pi f . \delta_i)}. \tilde{E_i}(f)\\
    \delta_i = \argmin_{\delta_i}\left\Vert T_{ref}-T_{i-corrected}(\delta_i)\right\Vert_2.\label{eq:opt1}
\end{gather}

The goal in Eq.\ref{eq:opt1} is to find for each time trace the drift which minimize the distance between the reference and the corrected time trace.

    
Fig.~\ref{figrda} shows the mean and the precision on $1\,000$ time traces after the drift correction. The signal to noise ratio increase from $28$dB to $55$dB, which means approximately a gain of $27$dB.

\begin{figure}[!htb]
   \centering
   \includegraphics[width=8cm]{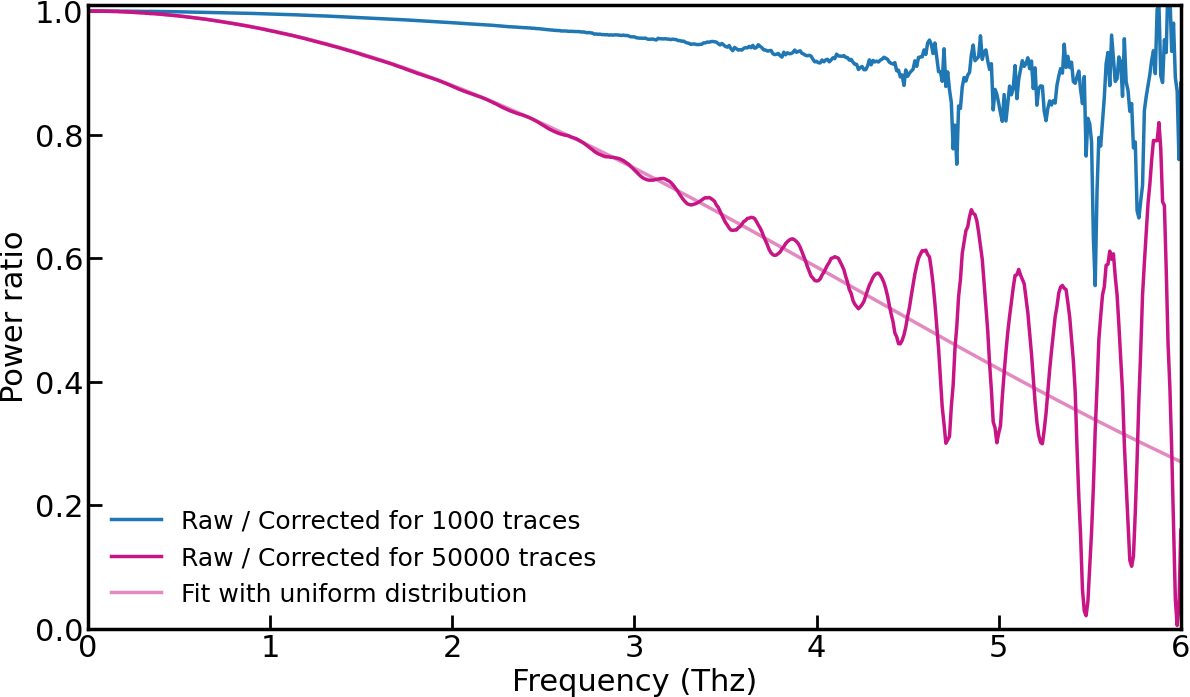}
   \caption{Power ratio of spectra obtained from a raw average and a corrected average showing the multiplicative bias caused by the drift. Shown for 1000 and 50000 time traces. The drift attenuation for 50000 time traces is fitted with the model from equation~\ref{eq:attCorr}.} 
              \label{figfilter}
    \end{figure}

Furthermore, averaging shifted time traces is equivalent to applying a convolution low-pass filter to the signal as explained in \cite{souders1990}. Assuming a linear drift during the acquisition for an overall drift $\Delta$, for an average over a large number of traces, the attenuation factor in \emph{amplitude} caused by the drift can be approximated in the frequency domain as 
\begin{equation}
	A_\Delta(\omega) = \frac{1}{\Delta}\int_{-\frac{\Delta}{2}}^{\frac{\Delta}{2}}e^{j\omega\delta}\mathrm d \delta = \text{sinc}\left(\frac{\omega\Delta}{2}\right). \label{eq:attCorr}
\end{equation}
More generally, the attenuation is the Fourier transform of the delay distribution among all the time traces. Equation~\ref{eq:attCorr} corresponds to the special case of an uniform distribution. It is important to note that the attenuation does not depends directly on the number of traces, but only on the delay distribution. Hence, the simple average of the raw time traces is a biased estimator of the signal, and does not approach the real signal for large number of time traces. In fact, the delay distribution is more likely to be broader for long acquisitions, such that the accuracy of the uncorrected average \emph{decreases} with the number of time traces. Fig.~\ref{figfilter} shows the ratio between the corrected average and the raw average for \num{1000} and \num{50000} time traces. In this example, the delay drift caused a severe attenuation in the raw average, and gets much worse for a larger number of averaged time traces. The attenuation closely match the form predicted in equation \ref{eq:attCorr} up to \SI{4}{THz}, despite a non linear drift (see supplementary document for details on the actual drift).

Figure~\ref{figfilter} highlights the potentially misleading nature of the error caused  by the delay drift. Indeed, despite the signal being affected by more than 20\% at \SI{3}{THz}, the signal does not appear more "noisy", and it is very difficult to distinguish this error from a real signal if only the raw average is taken into consideration. In fact, it could well be an explanation of the dispersion in the published results of THz spectral analysis of water 
using THz-TDS techniques (\cite{lavancier2021thesis,smolyanskaya2018} and references therein), where the strong absorption makes long accumulations necessary .

\subsection{Delay line speed variation}
At the view of the major gain obtained by the delay correction at the first order, we decided to go a step further. In fact, the variations in speed of the delay line due to the temperature or any other environmental fluctuation, has already been reported as in \cite{jahn2016influence}. Hence, there can be at the second order, noise due to the phenomena describe above. It leads to a dilatation in the time axis where the measured signal is $E(t+\alpha t)$ instead of  $E(t)$. A dilatation in the time domain implies a contraction in the frequency domain, which is a nonlinear transformation. To correct this, we stay in the time domain and use the Taylor formula at first order. 

\begin{gather}
    E_{i-corrected}(t) = E_i(t) - (\alpha_i t)E'_i(t)\\
    \alpha_i = \argmin_{\alpha_i} \left\Vert T_{ref}-T_{i-corrected}(\alpha_i)\right\Vert_2\label{eq:opt2}
\end{gather}

After the delay line speed correction, the gain is really small and there is almost no difference in the time and frequency domain, when applied to a single peak reference as we are doing here. However, we performed experiments outside of the scope of this paper and realized that  this correction becomes important when analyzing time traces containing echoes as the one coming from the recording of a solid state sample (the echoes come from the Fabry Perrot effect).

\subsection{Laser amplitude fluctuation}
The gain after the drift correction is important, however there is still noise remaining. This noise corresponds to other sources such as shot noise and fluctuations in laser power. We suppose this noise is proportional to the signal with a coefficient $a_i$ for each time trace $i$. Hence the corrected signal in time domain is:
\begin{gather}
    E_{i-corrected}(t) = (1-a_i)E_i(t)\\
    a_i = \argmin _{a_i} \left\Vert T_{ref}-T_{i-corrected}(a_i)\right\Vert_2\label{eq:opt3}
\end{gather}

Here the goal is the same as before, find for each time trace the coefficient which minimize the distance between the reference and the corrected time trace.

   \begin{figure}[!htb]
   \centering
   \includegraphics[width=8cm]{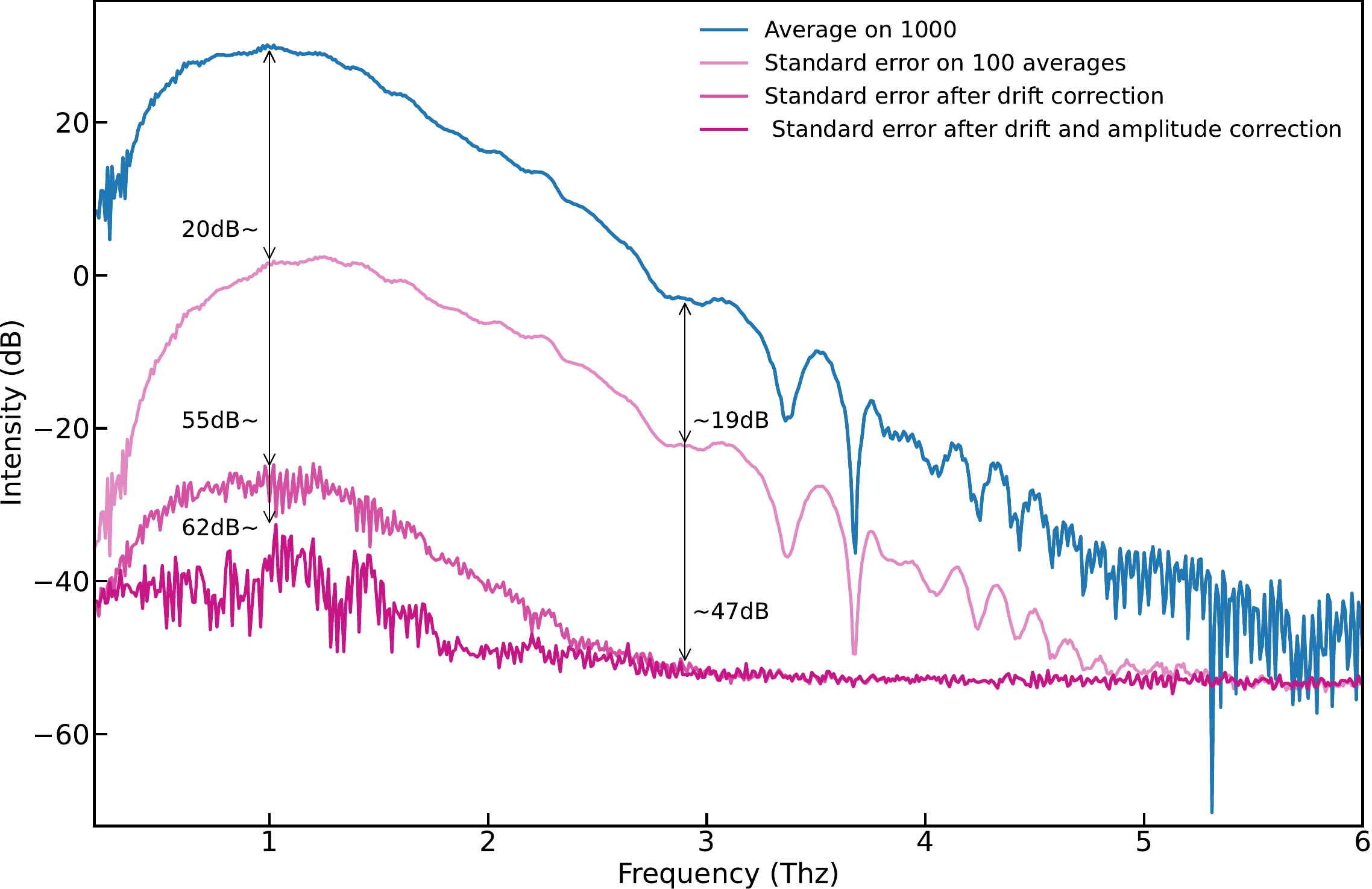}
   \caption{Mean and standard error in frequency and time domain after drift and amplitude correction}
              \label{figrda}
    \end{figure}

Fig.~\ref{figrda} shows the mean and the precision on \num{1000} time traces after the drift and the amplitude correction. In this case the signal estimated does not change. However the noise estimated does. Hence, the signal to noise ratio increase from \SI{55}{dB} to \SI{62}{dB}, which means approximately \SI{7}{dB} gain.

\subsection{Periodic sampling errors}
\begin{figure}[!htb]
   \centering
   \includegraphics[width=8cm]{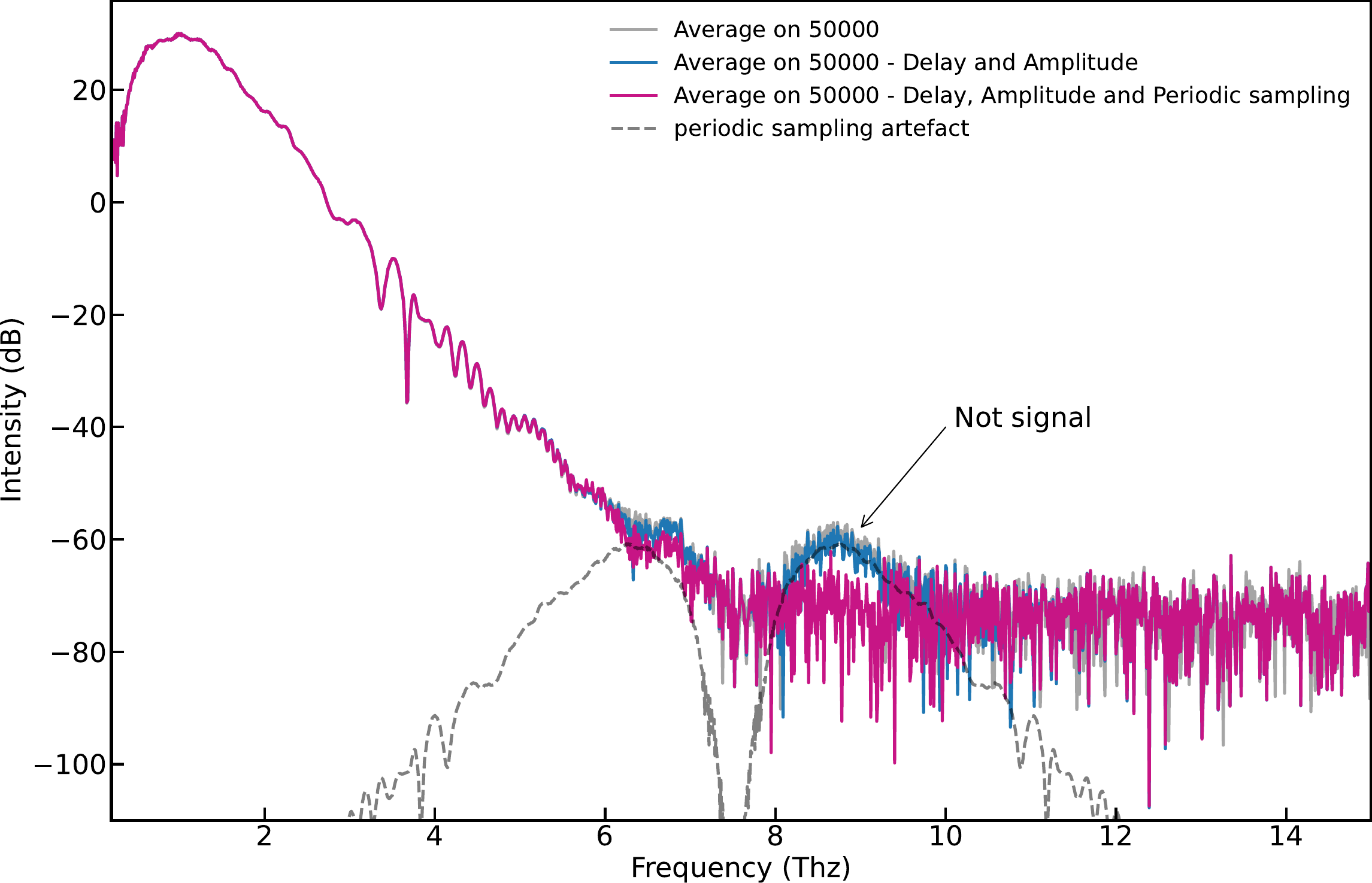}
   \caption{Spectrum with periodic sampling error beginning at \SI{7.5}{THz} before and after correction with a Tukey window ($\alpha = 0.05$)}
              \label{figps}
    \end{figure}

When averaging repeating measures, it appears that there is a distortion in the spectrum, which makes the spectrum of the estimated signal unusable and can significantly influence the extraction of the refractive index of the sample at high frequencies (typically $> \SI{7.5}{THz}$). Fig.~\ref{figps} shows the mean on \num{50000} time traces before correction in gray and after drift plus other corrections in blue. After approximately \SI{7.5}{THz} there is an artifact and between \SI{4.5}{THz} and \SI{7.5}{THz}, the artifact is present but hidden by the signal intensity, this area is then not reliable depending of the number of time traces accumulated. As explained in \cite{rehn2017periodic}, this distortion is called periodic sampling error and comes from periodic deviation of the delay line position. It leads to apparently higher artificial bandwidth due to parasitic mirror copies of the main pulse spectrum around the error’s frequency. Indeed, at each time $t$, instead of measuring the signal $E(t)$ the system measures $E(t+\sigma(t))$, $\sigma(t)$ being a periodic error in the time domain. It is possible to recover an approximation of the normal spectrum  from the measured spectrum, by minimizing the electric field in the concerned area. Indeed there is a frequency we call $f_{ps} \in \left\{f_0,\cdots,f_{\left\lfloor \frac{p-1}{2}\right\rfloor}\right\}$ that can be identify graphically, where the error start. The corrected signal is:

\begin{gather}
    E_{i-corrected}(t) = E_i(t) - E'_i(t) . \sigma_i(t) \\
    \sigma_i(t) = A_i\cos(\nu_i t + \phi_i) \\
    (A_i, \nu_i, \phi_i) = \argmin_{A_i, \nu_i, \phi_i} \sum_{f \in f_{ps},\cdots,f_{\left\lfloor \frac{p-1}{2}\right\rfloor}} |\tilde{E}_i(f)|
\end{gather}

$\sigma(t)$ is found by minimizing the integral of the spectrum in the concerned area.  $f_{ps}$ is approximately $7.5$THz in Fig.~\ref{figps} showing this noise and its reduction by the proposed method.

In this section, we proposed and implemented in Correct@TDS two important corrections for the signal estimator in THz-TDS. As in sampling oscilloscope for instance \cite{verbeyst2006}, the time sampling drift lead to low-pass effect on the amplitude characteristic that can be even more detrimental here because if its effect on the reference would be more important than on the sample this would lead to an artificial gain measurements on the sample. We anticipate that, as in fast electronics and ultrasonic systems\cite{gao2021}, further corrections in both the systems and in the data processing would further improve the signal estimator in the future. This will be even more interesting and impactful regarding the new developed THz-TDS ultra-rapid systems. There the mechanical delay line is replaced by optoelectronic elements as in asynchronous optical sampling systems (ASOPS) \cite{kliebisch2016,oeri2020}, or by two coupled laser as dual comb spectroscopy \cite{fu2021}. Despites their speed to acquire a spectrum and their high resolution these systems are facing issues to reach as good signal to noise ratio as the mechanical systems for the same integration time and will surely benefit of specific and more sophisticated noise correction and signal estimators. 

\section{Material parameters extraction from estimated signals}
Spectroscopy studies how a sample, a material, behaves when crossed by an electromagnetic wave. The goal is to extract parameters describing the physics of the electric charges in the sample. In THz-TDS, this extraction requires the knowledge of the THz pulse before and after the sample, which translate in without and with the sample. The measurements without the sample is called the reference pulse and the one with the sample the sample pulse. Once recorded both pulses are processed to estimate both the signal and the noise. 
\subsection{Fitting process}
Then to go further in the interpretation of the measurements the physicist select a permittivity model set in a transfer function transforming the reference pulse into the sample pulse. Since most of the dielectric function models are depicted in the frequency domain, most of the fits are performed there. However, several groups of which we are part have decided to perform it in the time domain to be as close as possible to the experimental signal. In \cite{peretti2018thz}, we implement our approach in a software called called Fit@TDS where the cost function to compare a model with its set of parameters $P$ to the experiments, is defined as the squared error between the sample pulse and the model fitted curve:  

\begin{gather}
    C_{model}\{P\} = \left\Vert E_{sample}-E_{model}\{P\})\right\Vert_2\\
    E_{model}\{P\} = f(P, E_{ref})
\end{gather}
The function $f$ corresponds to the model, while $C_{model}\{P\}$ is the error associated with the model and its parameters $P$.

Most fitting models implicitly assume that the noise is Gaussian and independent and identically distributed (i.i.d.). However, as shown in Fig.~\ref{figraw}, the noise present in THz-TDS signals is not a Gaussian white noise, then its impact cannot be ignored. To address this issue, \cite{mohtashemi2021maximum} proposed to incorporate the noise correlation matrix, into the cost function in the fitting process to account for measurement uncertainties. This approach involves normalizing the model cost with the matrix and selecting the best fitting model based on the Akaike criterion, which balances goodness of fit and model complexity in terms of number of parameters \cite{lavancier2024}.

The noise correlation matrix called here $M_{noise}$ contains the information about the noise in the sample signal, the reference signal and the correlations. It enables a proper weighting of the cost function during the fitting process and therefore  improves the accuracy of the material parameter extraction. It leads to a new associated cost function to the fitting model:

\begin{gather}
    C_{model}\{P\} = \left\Vert [M_{noise}]^{-1/2} (E_{sample}-E_{model}\{P\})\right\Vert_2 \label{eq2}\\
    M_{noise} = \mathlarger{\Sigma}_{E_{sample}} + \mathlarger{\Sigma}_{h * E_{ref}} \label{eq3}\\
    C_{AIC}\{P\} = C_{model}\{P\}  + 2N_{P} \label{eq4}
\end{gather}
$C_{AIC}$\{P\}  is the Akaike criterion associated with the model and its parameters ${P}$, $N_{P}$ is the number of parameters of the model.
The equation \eqref{eq3} represents the noise correlation matrix as a sum of two covariance matrices. The first term, $\mathlarger{\Sigma}_{E_{sample}}$, is the covariance matrix of the sample measurements. The second term, $ \mathlarger{\Sigma}_{h * E_{ref}}$, is the covariance matrix of the reference measurements, convolved with the transfer function $h$ of the THz-TDS system. The transfer function $h$ accounts for the distortion of the THz pulse as it propagates through the system. By convolving the reference with the transfer function, the noise correlation matrix of the reference is transferred to the sample one and added to the noise correlation matrix of the sample.

In practice, the true values of $h$ and the covariances $\mathlarger{\Sigma}_{E_{sample}}$ and $ \mathlarger{\Sigma}_{h * E_{ref}}$ are  unknown and have to be estimated from measured data. The estimation of the off-diagonal values of the covariance matrices is important because there are often correlations, as we have seen previously. Moreover, these correlations increase in the presence of a sample due to its own variations, making their estimation even more crucial. Therefore, the estimators should be able to approximate these off-diagonal values accurately. One approach is to estimate them from a set of repeated measurements, but there can be errors caused by the system, leading to a high variance in the signal. This problem highlights the importance of our corrections, which reduce the noise and enable a less biased estimation of the covariance matrices.

\subsection{Noise correlation matrix}
Covariance matrices are used in all field of signal from sampling oscilloscope \cite{humphreys2012}, acoustics \cite{yu2022} and biomedical application \cite{ali2021}. It serves, for instance, to propagate uncertainties, to shrink huge data to a reasonable size, and to obtain the necessary normalization \cite{humphreys2012} in a fitting process. In our case, we first estimate $E_{sample}$ and $E_{ref}$ with the proposed method : averaging $n_{acc}$ time traces after correcting the delay line and the laser amplitude errors.
\begin{gather}
    E_{ref} \approx E_{ref - n_{acc}} = \frac{1}{n_{acc}}\sum_{i = 1}^{n_{acc}} T_{i-\{ref\}}\\
   E_{sample} \approx E_{sample - n_{acc}} = \frac{1}{n_{acc}}\sum_{i = 1}^{n_{acc}} T_{i-\{sample\}}
\end{gather}

Following, the empirical value of the transfer function $h$ is derived as follow:
\begin{gather}
h = \text{TF}^{-1}\left[\frac{\tilde{E}_{sample}}{\tilde{E}_{ref}}\right] \approx \text{TF}^{-1}\left[\frac{\tilde{E}_{sample - n_{acc}}}{\tilde{E}_{ref - n_{acc}}}\right]
\end{gather}

Also, the computation of $M_{noise}$ requires to know the covariance matrices associated to the sample and the transformed reference. Since $E_{sample}$ and $h*E_{ref}$ are estimated by averaging $n_{acc}$ corrected time traces, their covariance matrices can be estimated using the following formula:

\begin{gather}
    \mathlarger{\Sigma}_{ h*E_{ref}} \approx  \frac{1}{n_{acc}}\mathlarger{\Sigma}_{ h*E_{ref- n_{acc}}}\\
        \mathlarger{\Sigma}_{ E_{sample}} \approx  \frac{1}{n_{acc}}\mathlarger{\Sigma}_{ E_{sample- n_{acc}}}
\end{gather}


We divide by $n_{acc}$ in the formula because we are estimating the covariance of the mean of the corrected time traces, not the covariance of each individual trace. When we average the $n_{acc}$ corrected time traces, we reduce the variance of the resulting mean by a factor of $n_{acc}$. Thus, dividing by $n_{acc}$ in the covariance computation accounts for this reduction in variance due to averaging.

Finally, the normalization in equation \eqref{eq2} implies the computation of the inverse of the noise correlation matrix. If the two covariance matrices in equation \eqref{eq3} are invertibles, that means definite positive then their sum is also invertible. But there is no assumption on this invertibility if the two covariance matrices are singulars. 

\subsection{Covariance and precision matrices estimators}

A necessary condition for the inversion of an estimated covariance is that the number of sample meaning the number of repeated measurements used for the estimation, must be strictly greater than the number of variables meaning the number of time sampling points. In practice, the acquisition of sufficient number of time traces could be very long and an unpractical when dealing with numerous experimental conditions to test. For example, a time trace of \SI{400}{ps} is about \num{12500} time sampling points, and the acquisition of \num{12501} time traces is about $2$ hours. Also, the more time traces are accumulated, the more drifts errors are accumulated from intrinsic instability.

The literature offers various methods to approximate well-conditioned covariance matrices. Our goal is to propose a simple, reproducible process for approximating well-conditioned covariance matrices and to integrate this process into our open-source software solution, Correct@TDS, making it easily accessible to users. To achieve this, we utilize methods implemented in the widely-used Python library, Scikit-learn \cite{pedregosa2011scikit}. We examine three covariance estimation methods: empirical covariance, shrinkage methods, and graphical lasso.


The empirical covariance matrix $\Sigma_{emp}$ , can be unreliable as an estimator when $n$ is close to or less than $p$, leading to ill-conditioned covariance estimates due to the large magnitude gap between the eigenvalues. To avoid this inversion problem, the shrinkage method has been proposed by \cite{ledoit2004well} (Ledoit-Wolf shrinkage). It is a transformation of the empirical covariance matrix toward a target matrix with regularization techniques:

\begin{gather}
\mathlarger{\Sigma}_{shrunk} = (1-\alpha) \mathlarger{\Sigma}_{emp} + \alpha \frac{\Tr(\mathlarger{\Sigma}_{emp})}{p}I
\end{gather}

The target matrix considered here is proportional to the weighted sum of the identity matrix and the empirical matrix. The optimal value of $\alpha$ is computed by finding a bias-variance trade-off with no assumption on the data distribution. Another methods called  Oracle Approximating Shrinkage method \cite{chen2010shrinkage} had been proposed as an improvement of Ledoit-Wolf shrinkage where $\alpha$ is computed under Gaussian distribution assumption on the samples.

There are also methods that rely on solving the covariance selection problem to approximate sparse inverse covariance matrices with lasso regularization \cite{meinshausen2006high} \cite{yuan2007model}  \cite{friedman2008sparse}. These methods have the advantage to be able to recover the off diagonal structure of the covariance matrix. Since two independent variables will have zero coefficient in the precision matrix, the idea is to estimate a sparse covariance matrix of multivariate Gaussian distributed observations by minimizing the penalized negative log-likelihood:

\begin{gather}
\mathlarger{\Sigma}_{glasso} = \argmin_{\mathlarger{\Sigma}} -\log\left(\det \mathlarger{\Sigma}\right) + \Tr\left(\mathlarger{\Sigma}_{emp}\mathlarger{\Sigma}\right)  + \alpha \left\Vert\mathlarger{\Sigma}\right\Vert_1
\end{gather}

The coefficient $\alpha$ is the lasso penalty; its value is found by comparing different values on a grid of $\alpha$ values. The larger alpha is, the more sparse the matrix is. This method tends to work better when $n\leq p$, however the lasso penalization based methods can be inconsistent on some graph structures \cite{heinavaara2016inconsistency}. In \cite{friedman2008sparse}, they proposed the graphical lasso, a fast algorithm based on coordinate descent approach to solve this problem.

For the sake of the simplicity in the explanations, we limited the scope of the paper to data of a reference, meaning without samples in the estimation of $\mathlarger{\Sigma}_{ E_{ref-n_{acc}}}$ and its inverse. It worth noticing that the process is the same for $\mathlarger{\Sigma}_{ E_{sample-n_{acc}}}$ and $\mathlarger{\Sigma}_{ h*E_{ref-n_{acc}}}$ and we expect that the noise induced by the signal will be lower simply because a sample will reduce the magnitude of the signal. 

The tests are done on the measured time traces on \SI{100}{ps} in the third experiment in section 2. Since we don't know the distribution of the residual noise in Fig.~\ref{figrda}, we tried different estimation methods for different values of $n$ with $n\leq p$ and compared them to targets matrices which are supposed to be the true covariance/precision matrices. Those targets matrices are obtained by computing the empirical covariance and its inverse on  $n = \num{100000}$ time traces with $p = \num{3000}$ and $p/n = 0.03$. We assume that we have enough data for a well conditioned matrix close to the true matrices.

\subsection{Results and discussion}
Our goal was to compare the different methods to estimate the covariance and precision matrices. On this basis, we tested Ledoit Wolf shrinkage, Oracle Approximating shrinkage and the Graphical Lasso. This section shows the obtained results.

\subsubsection{Covariance matrix}

    \begin{figure}[!htb]
    \centering
    \includegraphics[width=8cm]{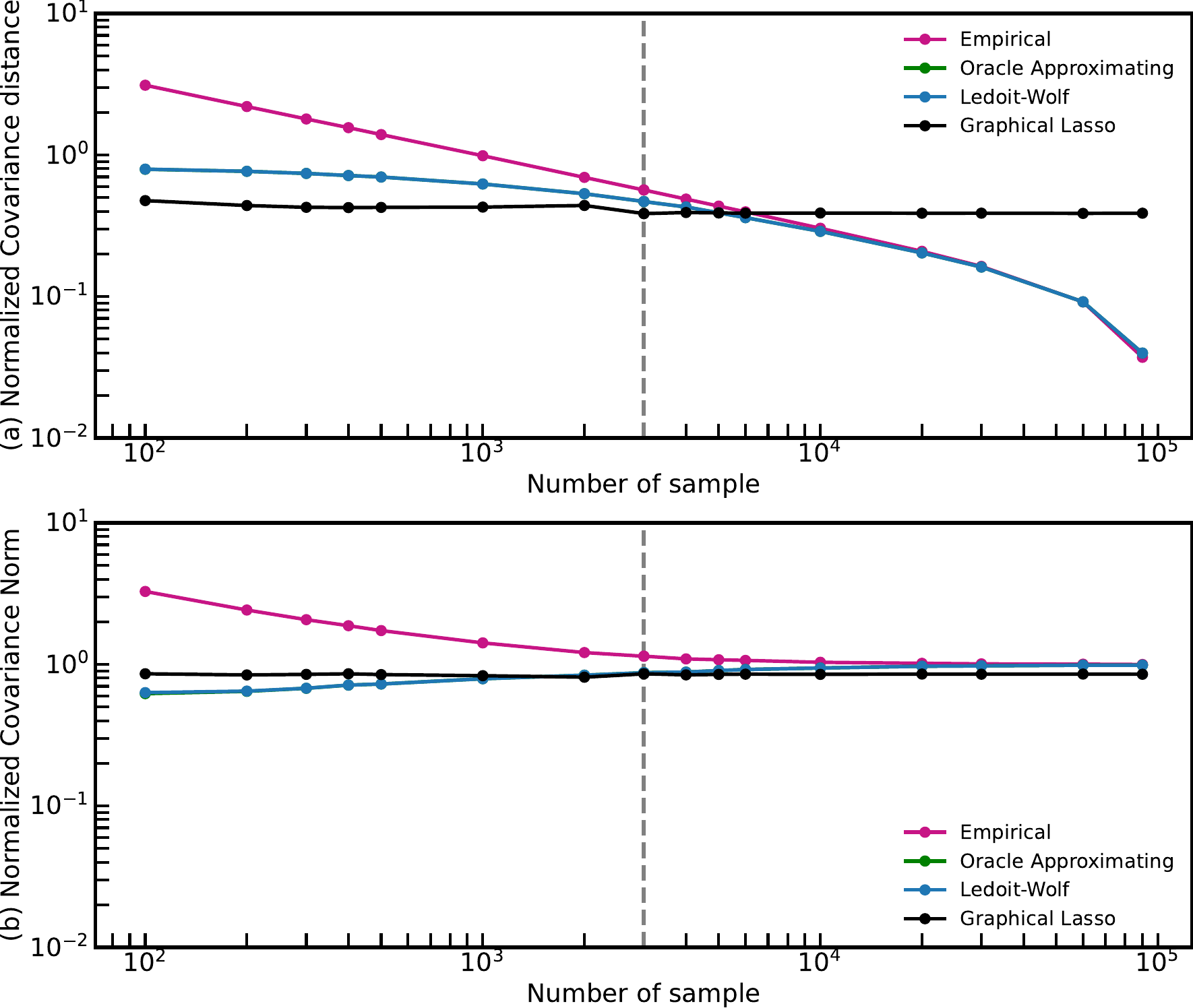} 
    \caption{Comparison between Ledoit Wolf shrinkage, Oracle Approximating shrinkage and the Graphical Lasso methods for the covariance matrix estimation.
    (a) Normalized distance between estimated and target covariance matrices for different number of sample showing a good estimation with the graphical lasso already with 100 time traces.
    (b) Normalized norm of the estimated covariance matrices showing that the norm of the estimated matrices are in the range of the targeted ones and that the graphical Lasso performs better
    }\label{figcov_metric}
\end{figure}

Figure~\ref{figcov_metric}(a) illustrates the normalized distance between the estimated covariance matrix for different number of sample and the empirical target covariance matrix. This distance is defined as the Frobenius norm of the difference between the two matrices, normalized by the Frobenius norm of the target matrix, with an expected value of 0. The gray line indicates the point at which $n=p+1$, indicating the invertibility of the covariance matrices. We can see that the two shrinkage methods produce similar results despite their different assumptions about the Gaussian distribution. When $n \leq p$, which is the relevant scenario of interest for us, the graphical lasso algorithm performs better even with small number of time traces such as $100$.

In addition to the comparison of the distances, we also compare the norm of the estimated matrices to the norm of the target matrices. Figure ~\ref{figcov_metric}(b) shows the ratio between these two norms, with an expected value of 1. The ratio ranges from approximately $0.6$ to $1$ depending on the estimation method. These results suggest that the estimated matrices are in close neighborhood of the target matrix and have similar norms, which indicates that the estimation methods produce accurate results.

Moreover, in order to evaluate the accuracy of the estimators in retrieving the covariance matrix structure and correlation values, a plot was generated.

\begin{figure*}[!htb]
    \centering
    \subfloat[\centering Empirical before correction (with filter at low frequencies)]{{\includegraphics[width=8cm]{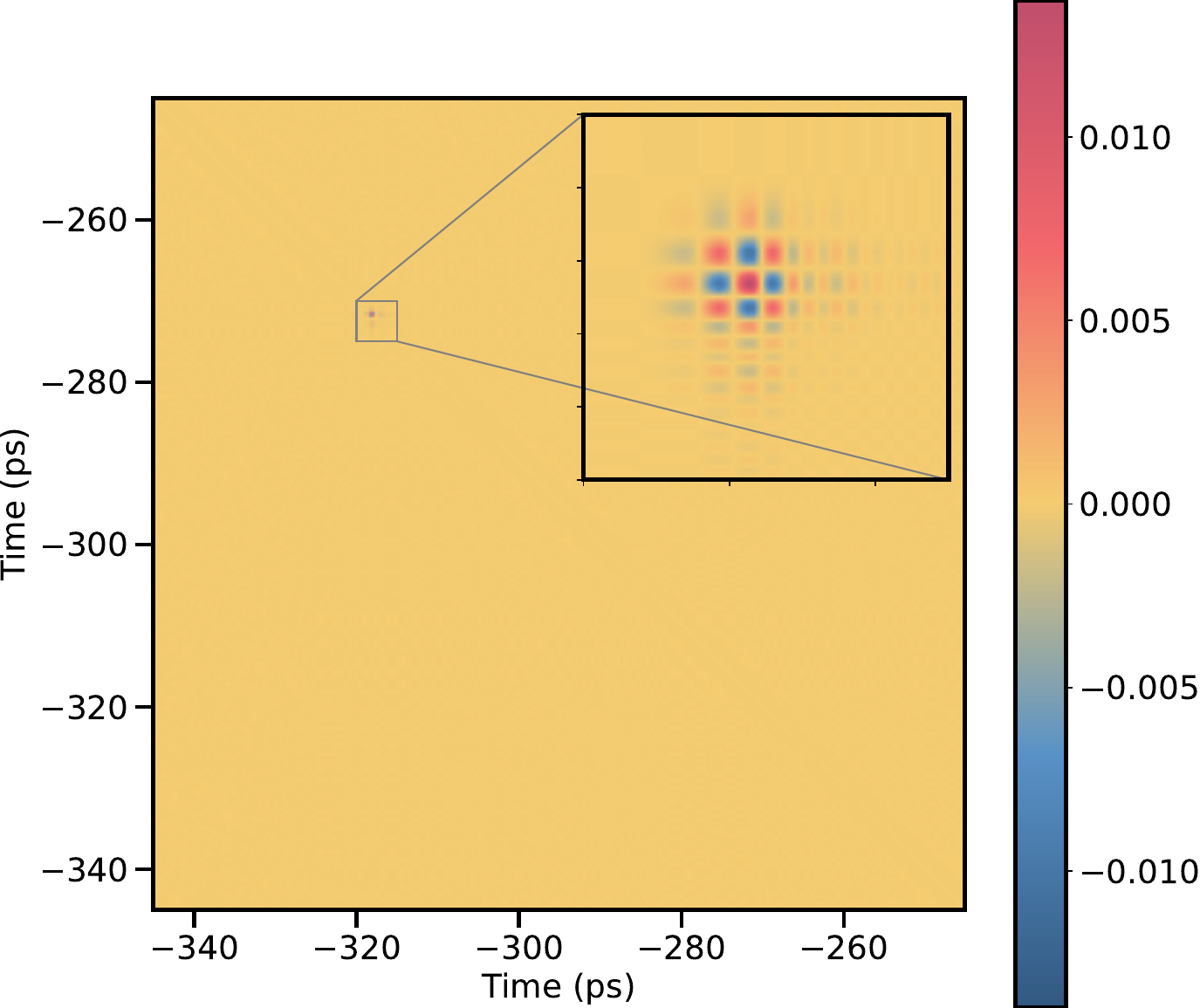} }}%
    \subfloat[\centering Empirical after correction]{{\includegraphics[width=8cm]{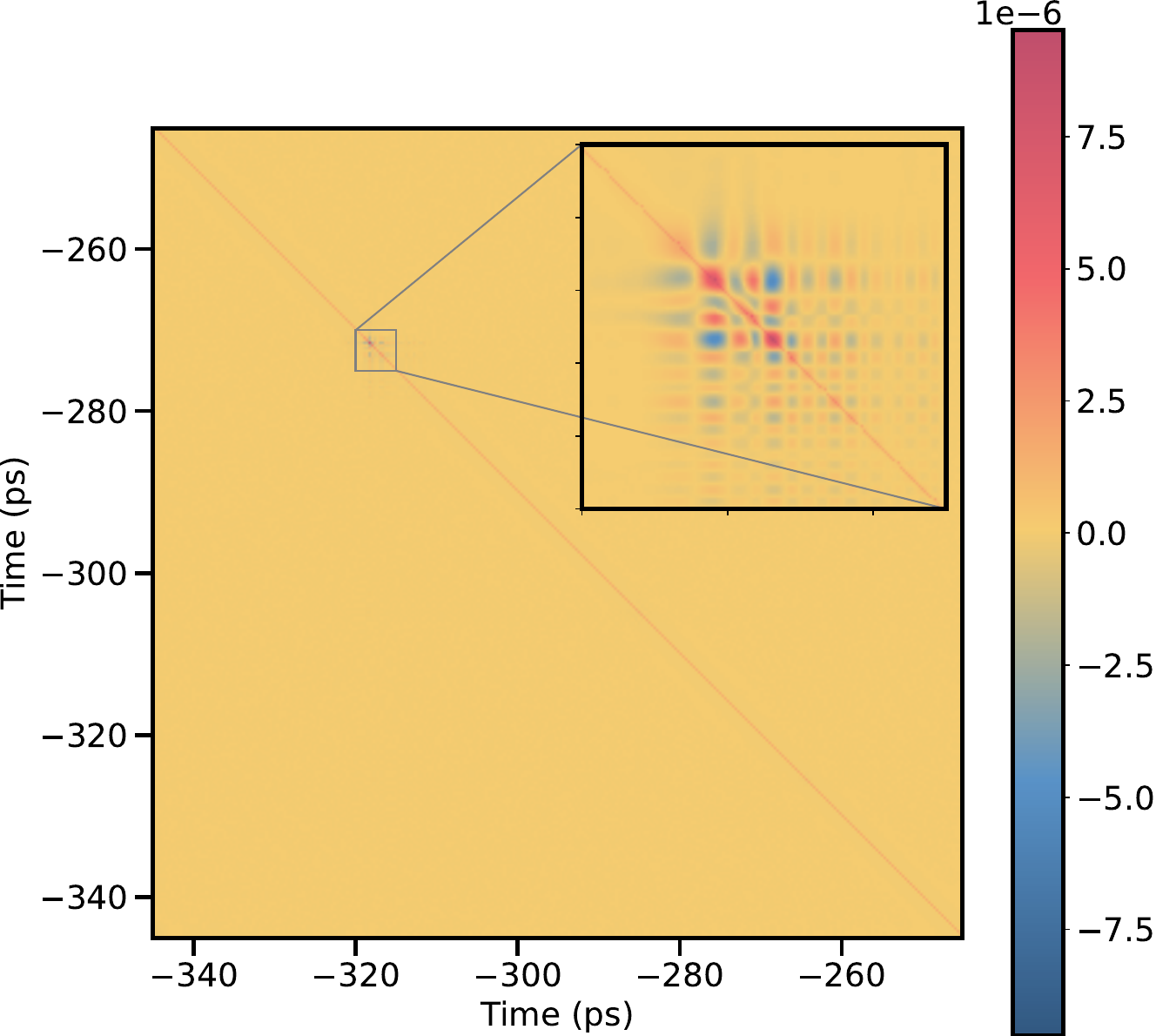} }}%
    \caption{Empirical covariance before correction and after correction for \num{100000} time traces showing the importance of the correction to get a covariance immune to drift}
    \label{figcov1}
\end{figure*}

    \begin{figure*}[!htb]
    \centering
    \subfloat[\centering Graphical Lasso]{{\includegraphics[width=8cm]{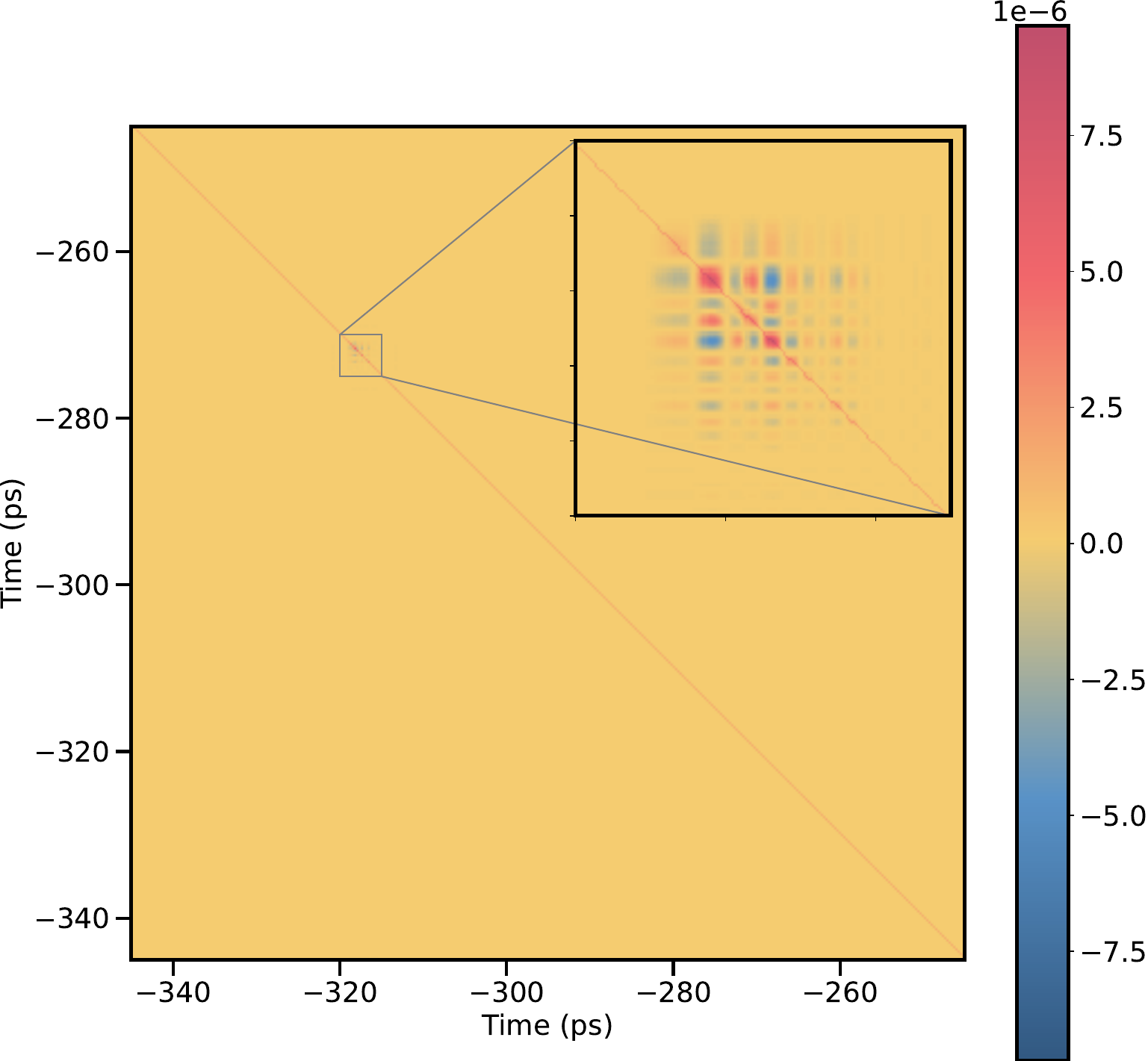} }}%
    \subfloat[\centering Ledoit-Wolf]{{\includegraphics[width=8cm]{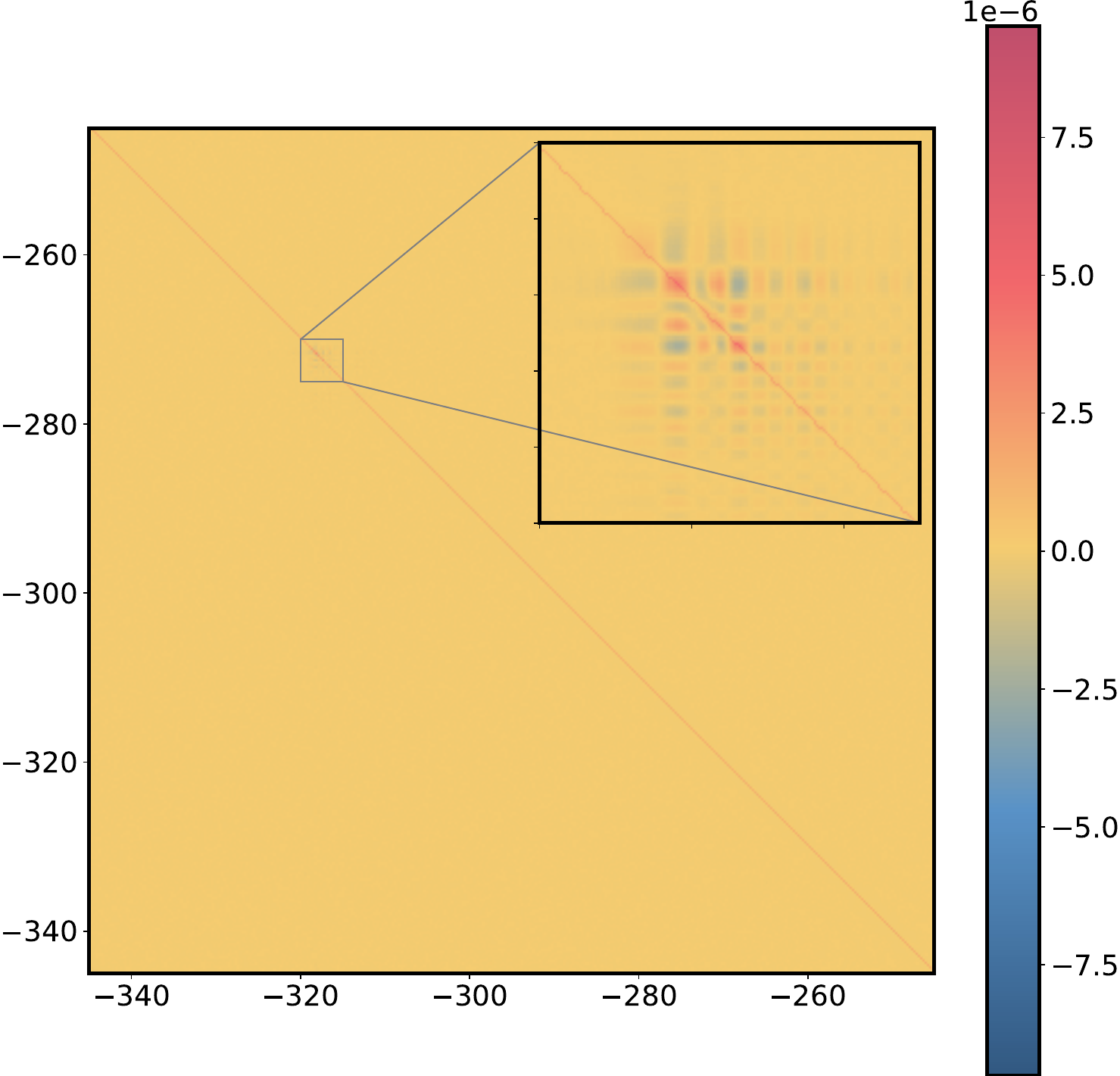} }}%
    \caption{Estimated Covariance with Graphical Lasso and Ledoit-Wolf algorithms for \num{1000} time traces}
    \label{figcov2}
\end{figure*}

Figure~\ref{figcov1}(a) displays the empirical covariance matrix before correction on \num{100000} time traces. Before correction, the diagonal values are not large compared to the off-diagonal values, indicating a certain degree of correlation between the variables that represent each time point in the signal. The correlations are certainly due to noise or artifacts in the signal and need to be removed or corrected. It should be noted that here, a low frequencies filter (refer section 2) is applied, otherwise the correlations would be stronger. Furthermore, the highest positive and negative covariance values are close to the main pulse, reflecting the position of the main pulse. 

In Figure~\ref{figcov1}(b), the empirical covariance matrix after correction on \num{100000} time traces is shown as the target matrix. The correction leads to a reduction of covariance values by three orders of magnitude, and a change in the diagonal elements, which become larger than the off-diagonal elements. Moreover, the correlation due to the main pulse is still reflected. This suggests that the correlation due to noise or unwanted artifacts in the signal are corrected. Hence we can conclude that the corrections help to obtain an almost accurate representation of the underlying signal.

Figure~\ref{figcov2} illustrates the estimated covariance matrix with $n_{acc}=1000$ and $p=3000$ with Graphical Lasso (a) and Ledoit-Wolf (b). They have the same structure with the target variables, meaning they refleted the main pulse correlation and the large diagonal values. However, the Graphical Lasso algorithm appears to provide a more precise estimation of the main pulse correlation compared to the target covariance matrix, which is consistent with the results shown in Figure~\ref{figcov_metric}(a) where the graphical lasso algorithm performed better in terms of distance to the target matrix.

\subsubsection{Precision matrix}

The most challenging task is now to determine which estimator is the most precise for retrieving the precision matrix. Obviously we have a curse of dimensionality problem and there is no optimal solution for this. The same comparisons as for the covariance estimator are done for the precision.

    \begin{figure}[!htb]
    \centering
    \includegraphics[width=8cm]{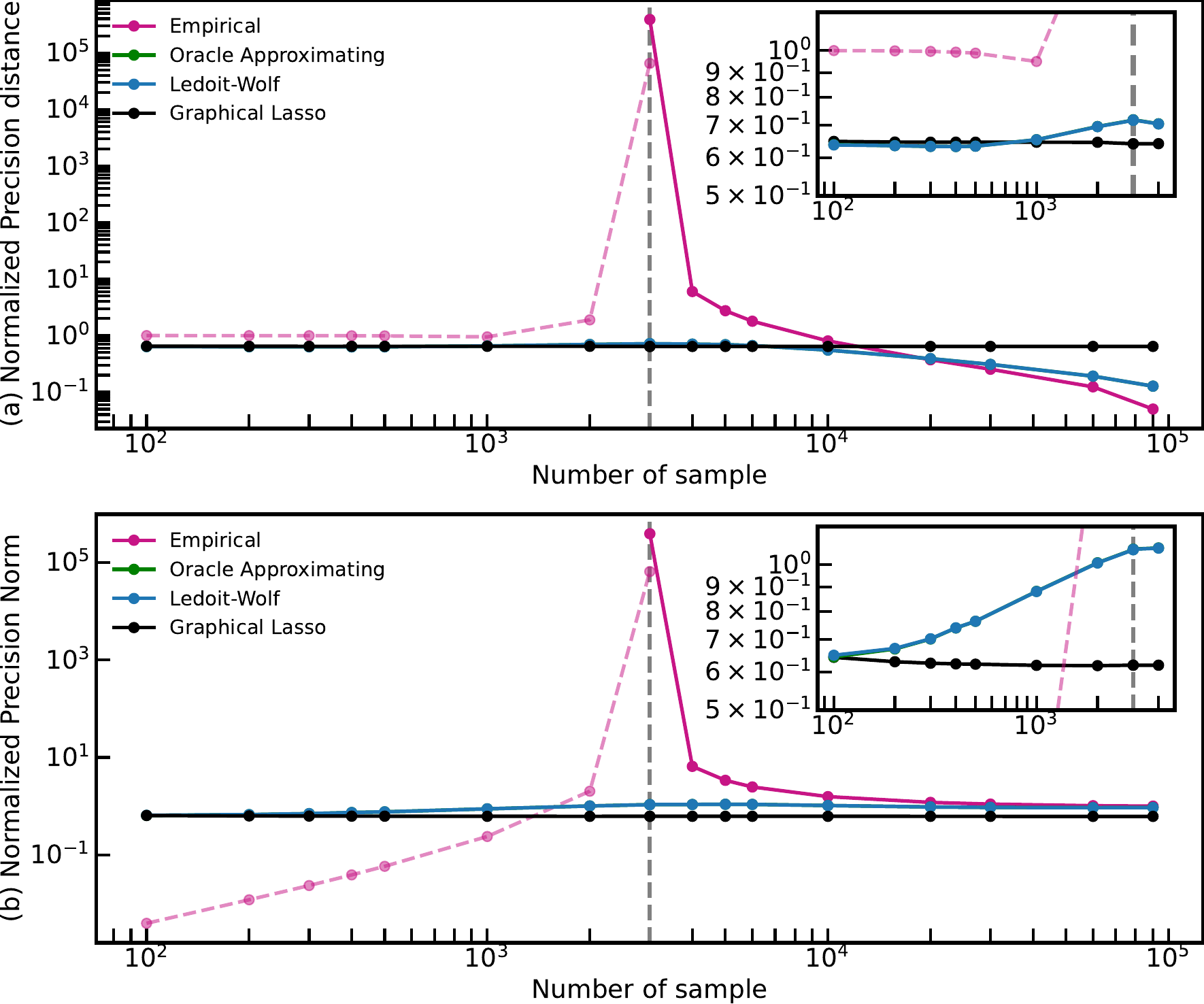} 
    \caption{Comparison between Ledoit Wolf shrinkage, Oracle Approximating shrinkage and the Graphical Lasso methods for the precision matrix estimation.
    (a) Normalized distance between estimated and target precision matrices for different number of sample showing almost the same results for the three methods when $n\leq p$.
    (b) Normalized norm of the estimated precision matrices showing that the norm of the estimated matrices are in the range of the targeted ones.
    }\label{figprec_metric}
\end{figure}

Figure~\ref{figprec_metric}(a) shows the normalized distances between the estimated precision matrices and the target precision matrix (inverse of the empirical target covariance matrix). The two shrinkage methods also produce here similar results. In the case where $n \leq p$, where empirical covariance matrix is not invertible, the distances for all methods are almost in the same order of magnitude. However when we get closer to the invertibility line, the graphical lasso seems to works better.

Regarding the normalized norms of the estimated matrices in Figure~\ref{figprec_metric}(b), we can observe that, the norms with the shrinkage methods get closer to the target with more samples while the norms with graphical lasso are almost constant. However all values also range from 0.6 to 1 depending on the estimation method indicating that the estimated matrices are also in close neighborhood of the target matrix.

\begin{figure*}[!htb]
    \centering
    \subfloat[\centering Empirical before correction (with filter at low frequencies)]{{\includegraphics[width=8cm]{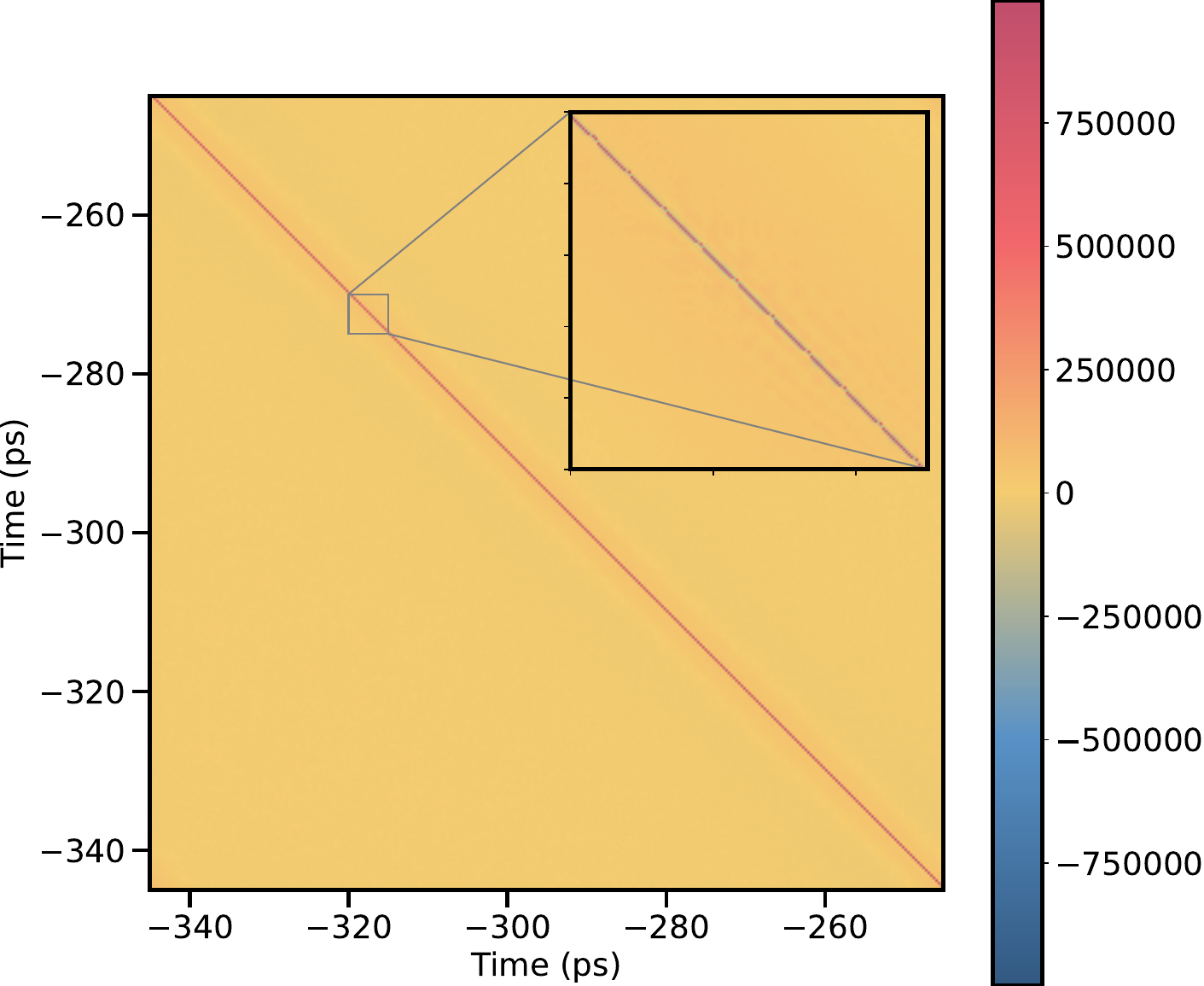} }}%
    \subfloat[\centering Empirical after correction]{{\includegraphics[width=7.5cm]{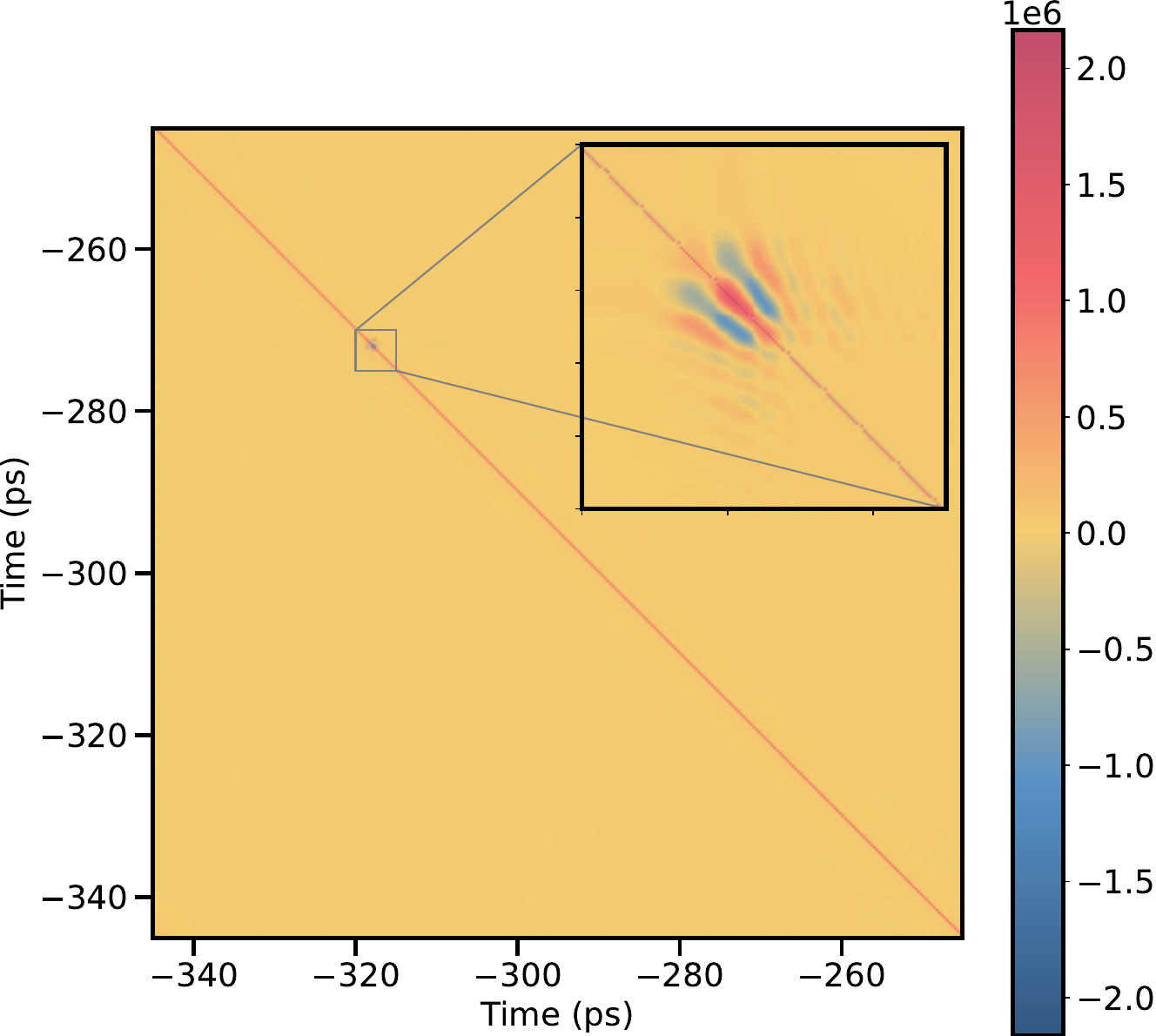} }}%
    \caption{Empirical Precision before correction and after correction for \num{100000} time traces}
    \label{figprec1}
\end{figure*}

\begin{figure*}[!htb]
    \centering
    \subfloat[\centering Graphical Lasso]{{\includegraphics[width=8cm]{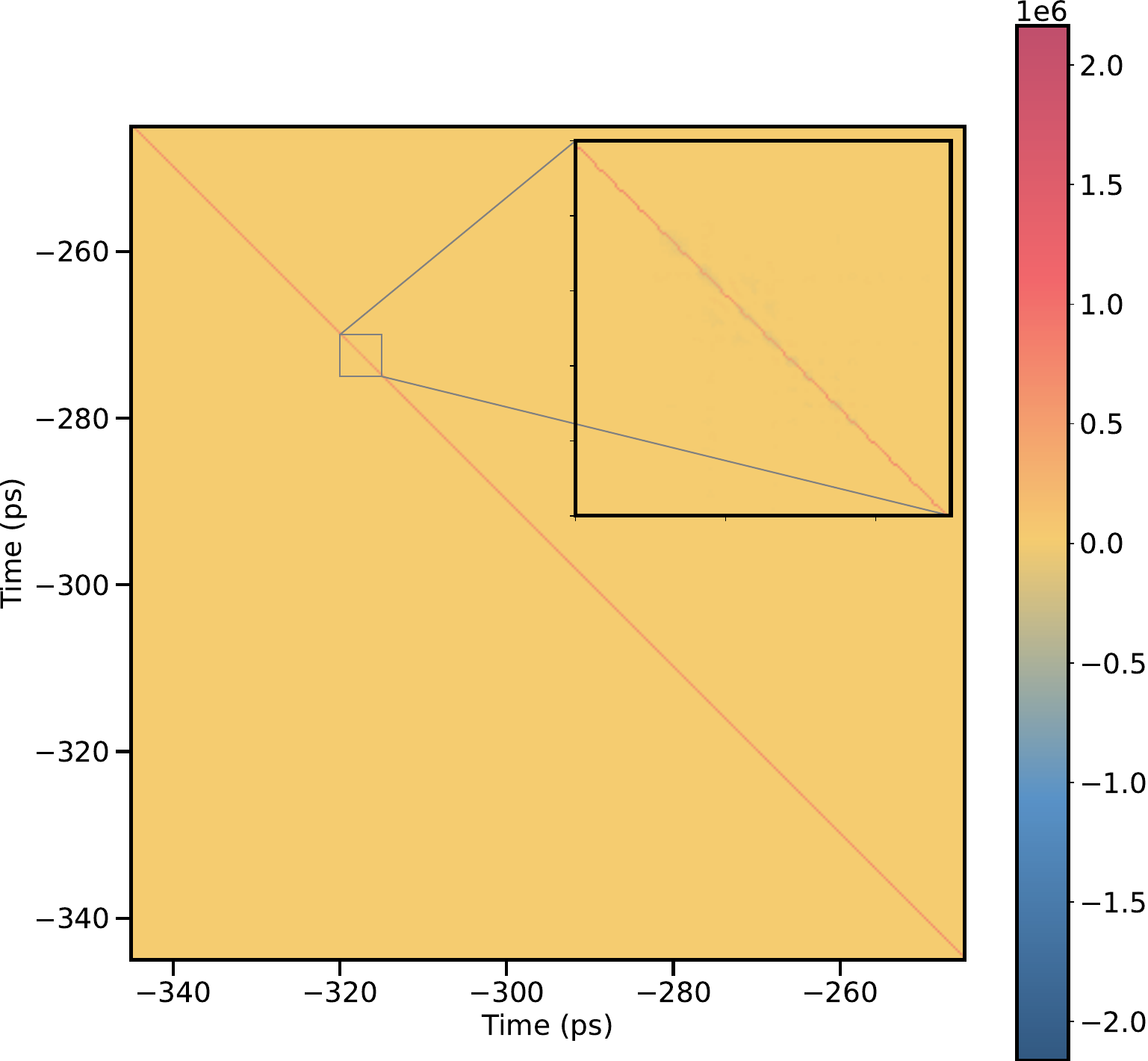} }}%
    \subfloat[\centering Ledoit-Wolf]{{\includegraphics[width=7.5cm]{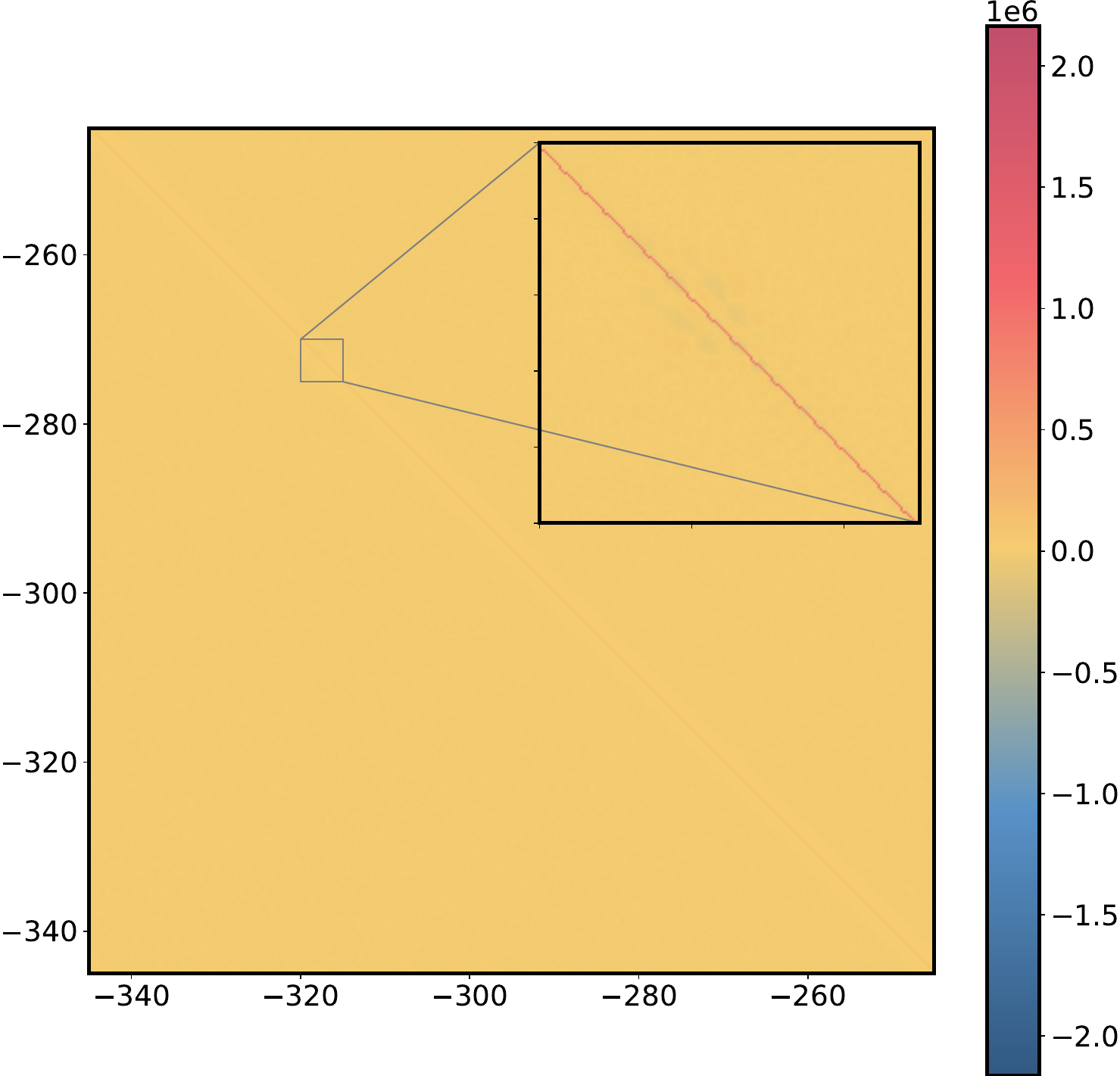} }}%

    \caption{Estimated Precision with Graphical Lasso and Ledoit-Wolf algorithms for \num{1\,000} time traces}
    \label{figprec2}
\end{figure*}

In Fig.~\ref{figprec1}, the precision matrix (inverse of the empirical covariance matrix) is shown before and after correction on \num{100000} time traces. Despite the number of sample supposed sufficient used for computation, the precision matrix before correction does not reflect the strong correlation on the main pulse as in the covariance, which is contrary to the expectation that two independent variables would have a zero coefficient in the precision matrix, and vice versa. On the other hand, the precision matrix which is our target precision after correction appears to better capture the expected structure. Also their is an increase of the partial correlation values by one order of magnitude after correction.

In Fig.~\ref{figprec2}, the precision matrices obtained by applying the graphical lasso and Ledoit-Wolf methods on \num{1000} time traces are presented. Although these matrices appear to be closer to the target matrix (also referring to Fig.~\ref{figprec_metric}), the graphical lasso fails to capture the strong correlation on the pulse in the precision matrix. Still, this structure can be explained by the sparsity hypothesis of the precision matrix made in lasso algorithm, can affect the structure and make it different from the covariance matrix. As for Ledoit-Wolf, we observe a partial retrieval of the strong correlation, but it remains difficult for the estimators to capture it entirely. It can just be due to the fact that $1\,000$ time traces is not sufficient for this method.

\subsubsection{Best estimator}

Finally, graphical lasso seems to be the best estimator for our problem. The model's goal is to retrieve the matrix structure. It regularizes the empirical matrix and give sparse precision matrices. While the resulting matrices may not be entirely accurate, they are still suitable for our purpose that is to ensure a proper fit normalization taking into account the noise distribution and therefore giving better evaluation of the extracted physical parameters and not to get the exact values.  Additionally,  Graphical lasso has the advantage to give sparse matrix that will be advantageous in terms of computational costs during the fitting process.

Noise correlation matrix estimation is again a general problem in data processing where specific solution can be found for specific problems. Here, we took the first steps in their evaluation in the case of  THz-TDS with the goal of ensuring a proper weighting during the fit process giving the parameters interesting for the users of the setup. It leads to the first evaluation of these matrix for this setup to our knowledge with a reasonable accuracy using only python public libraries. We have no doubt that more sophisticated methods based on the recent works on the subject \cite{ollila2020,wei2023,ollila2022} will give quicker and more accurate evaluation of the matrix. Similarly to what we wrote previously, new systems based on optoelectronics or dual laser are able to record thousands of time traces per second and will benefit a lot from proper sparse evaluation of the noise correlation matrix to avoid to record huge volume of correlated data.

\section{Conclusion}
In conclusion, we have introduced a nonlinear signal estimator and several techniques for retrieving the noise correlation matrix for THz-TDS experiments. These methods have been implemented in our open-source software, Correct@TDS, readily available to the community. Our signal correction significantly improves the signal-to-noise ratio(over \SI{30}{dB} enhancement at \SI{1}{THz} for \num{1000} accumulated traces). This advancement transforms THz-TDS setups worldwide from capable of measuring small or very attenuated signals to excelling at measuring tiny variations of signal, such as in very thin samples or samples with subtle physical or chemical differences. Additionally, our work brought THz-TDS material parameter extraction to more reliable standards by incorporating the proper noise term as a weighting factor.

Our work opens several exciting perspectives in several areas. This work motivates experimentalists to search for hidden noise sources to further reduce noise even more and approach true Gaussian white noise. We are already beginning the integration of Correct@TDS with our existing software, FIT@TDS \cite{peretti2018thz}, to extract error bars for fitted parameters, even with our super-resolution technique \cite{eliet2022}.

Moreover, the fields of THz-TDS signal processing \cite{withayachumnankul2014,Mittleman1998} and parameter retrieval \cite{duvillaret1996reliable,dorney2001material,pupeza2007highly}  have a rich history of over two decades, primarily led by physicists with a focus on other areas of expertise. We anticipate our work to foster stronger collaborations between the THz and the signal processing communities. The latter could leverage concepts and techniques from speech recognition, seismology, radar, or biomedical echography to build faster, more accurate, and improved data processing methods. The widespread adoption of THz analysis in analytical fields like pharmaceuticals and nondestructive testing relies on continued improvements in measurement reliability and reproducibility, areas where standardized data processing can make significant contributions.

\section*{Acknowledgments}
      The authors thanks Claire Mantel for the fruitful discussions, the Région "Hauts-de-France" for the funding of the Diagnotera startair project and the DEUS MAROONER stimule project.

\vfill


\bibliographystyle{IEEEtranN}

\bibliography{biblio.bib}
\end{document}